# Intelligent Search of Correlated Alarms for GSM Networks with Model-based Constraints*

Qingguo Zheng Ke Xu Weifeng Lv Shilong Ma
National Lab of Software Development Environment
Department of Computer Science and Engineer
Beijing University of Aeronautics and Astronautics, Beijing 100083
{zqg, kexu, lwf , slma}@nlsde.buaa.edu.cn

## Abstract

In order to control the process of data mining and focus on the things of interest to us, many kinds of constraints have been added into the algorithms of data mining. However, discovering the correlated alarms in the alarm database needs deep domain constraints. Because the correlated alarms greatly depend on the logical and physical architecture of networks. Thus we use the network model as the constraints of algorithms, including *Scope constraint, Inter-correlated constraint* and *Intra-correlated constraint*, in our proposed algorithm called SMC (Search with Model-based Constraints). The experiments show that the SMC algorithm with Inter-correlated or Intra-correlated constraint is about two times faster than the algorithm with no constraints.

## 1. Introduction

Alarm correlation [1] plays an important role in isolating the root cause of problems from a stream of alarms for networks. Modern telecommunication networks generate a massive number of alarms, which make the analysis of alarm correlation more difficult than before. Many researchers look for a new method to solve the problem.

Mannila et al. [2] first studied the methods of mining frequent episodes in alarm sequences, which has been applied into TASA system [3]. Subsequently, many researchers [4,5] have studied adopting data mining methods [2,6,7,8] to make analysis of alarm correlation. However, the methods of data mining for alarm correlation will generate too many rules, which still need network administrator to distinguish the useful rules from the results of data mining. In order to control the process of data mining, make the process of data mining more focus, and reduce the number of rules being generated, many researchers [9,10,11] have studied putting various constraints, including Boolean expressions, regular expressions, and aggregation constraints etc., into the methods of mining association rules. But the analysis of alarm correlation will need deep domain constraints into the method of data mining, for the rules of alarm correlation are closely related to the logical and physical architecture of networks.

Model-Based Reasoning (MBR) was originally proposed in [12], the principles of which have been widely applied in the analysis of alarm correlation [1,13,14,15,16,17]. MBR system consists of a network model and a function model. In the case of telecommunication networks, network model is composed of network topology and network elements.

In this paper, we propose the intelligent search of correlated alarms with model-based constraints, called SMC (Search with Model-based Constraints), which use network configuration model as constraints to discover alarm correlation rules from the alarm database. We adopt three types of constraints, i.e. Scope constraint, Intra-correlated constraint and Inter-correlated constraint, according the network model of GSM networks. The experiments in this paper show that the algorithm of Inter-correlated constraint is faster than the algorithm of Intra-correlated constraint. The algorithms of Inter-correlated constraint and Intra-correlated constraint are about two times faster than the algorithm with no constraints.

The rest of this paper is organized as follows. In Section 2 we give the related works in the network management. In Section 3 we first briefly describe the architecture of GSM networks and then give the network configuration of GSM networks and define the Scope, intra-correlated alarms and inter-correlated alarms. Section 4 gives the definitions of the algorithm of searching correlated alarms. In Section 5

* This research was supported by National 973 Project of China Grant No. G1999032709 and No.G1999032701

we propose the intelligent search of correlated alarms with model-based constraints, called SMC (Searching based on Model constraints). The mode-based constraints include Scope constraint, Inter-correlated constraint and Intra-correlated constraint. In Section 5 experiments show that the SMC algorithm with Inter-correlated or Intra-correlated constraint is about two times faster than that with no constraints. Finally, we make conclusions in Section 6.

# 2. Related work

In the past, the principles of model-based reasoning [12] have been widely used in network management [1,13,14,15,16,17]. Fröhlich and Nejdl [13] introduced a model-based solution to the problem of alarm correlation for cellular phone networks (GSM). They proposed an abstract simulation to identify the cause of the propagation of alarm messages through the network by comparing the predicted behavior with the actual alarm pattern. It is very difficult to build the relevant functional, causal and structural semantics of model of GSM networks, because the system of GSM networks are very complex, including the wireless networks and fixed networks, and GSM networks evolve quickly, which now migrate into 2.5 Generation i.e. GPRS. So the SMC algorithm uses the network model as constraints to discover the correlated alarms from the database. Furthermore, the SMC algorithm adopts the concept of GSM networks rather than the network configuration as constraints and so avoids the problem of frequent changes of GSM networks.

Sabin et al. [16] proposed a constraint-based model and a problem-solving approach to network fault management. The approach is based on the concept of constraints which represent relations among network components. Subsequently, Sabin et al. [17] applied an extension to the constraint satisfaction paradigm, called composite constraint satisfaction, to facilitate modeling of complex systems and demonstrate the applicability of their approach on an example of a basic groupware service, namely, distributed database replication. The constraints of mode-based algorithms are only obtained from the expert, which is a bottleneck to the constraints-based model methods.

Many researchers of data mining also studied adding some constraints to the algorithms of data mining [9,10,11]. Srikant et al. [9] first studied the problem of discovering association rules in the present of constraints that are Boolean expressions over the presence of absence of items. Integrating such constraints as a post-processing step can greatly reduce the execution time. Raymond et al. [10] introduced and analyzed two properties: anti-monotonicity and succinctness, and developed characterizations of various constraints into four categories. After that they proposed a Mining algorithm called CAP, which achieves a maximized degree of pruning for all categories of constraints. Garofalakis et al. [11] proposed the use of Regular Expressions as a flexible constraint specification tool that enables user-controlled focus to be incorporated into the pattern mining process, and developed a family of novel algorithms called SPRIIT (sequential pattern mining with regular expression constraints) for mining frequent sequential patterns under constrains. These constraints above are too general to be fit for the analysis of alarm correlation in the network management. So some deep constraints of telecommunications networks are needed to discover the correlated alarms.

Some groups of network management also studied the layered model of networks [18,19,20]. Malheiros and Marcos Silva [18] presented a proposal of a general model for telecommunication network based on distributing sets of sub-networks into layers, according to functional affinity criteria. The proposed model uses an acyclic directed graph to represent a telecommunication network and is being used in network management applications, especially in alarm correlation. Appleby et al. [19] have developed a mode-based event correlation engine, called Yemanja, for multi-layer fault diagnosis. Yemanja uses the entity-models to encapsulate each device and conceptual layer. The entity-models contain a set of scenarios that embody its problem behavior. Barros et al. [20] presented a diagnostic approach based on a complete representation of the network in its logical and physical levels, which have been called the network *Configuration Model*. The Layered model is fit for the changes of networks, which is an overview from the whole network. So we use the concept network configuration model of the GSM network as the constraints of SMC (Search with Model-based constraints) algorithm.

# 3. Problem definition

## 3.1 The GSM networks

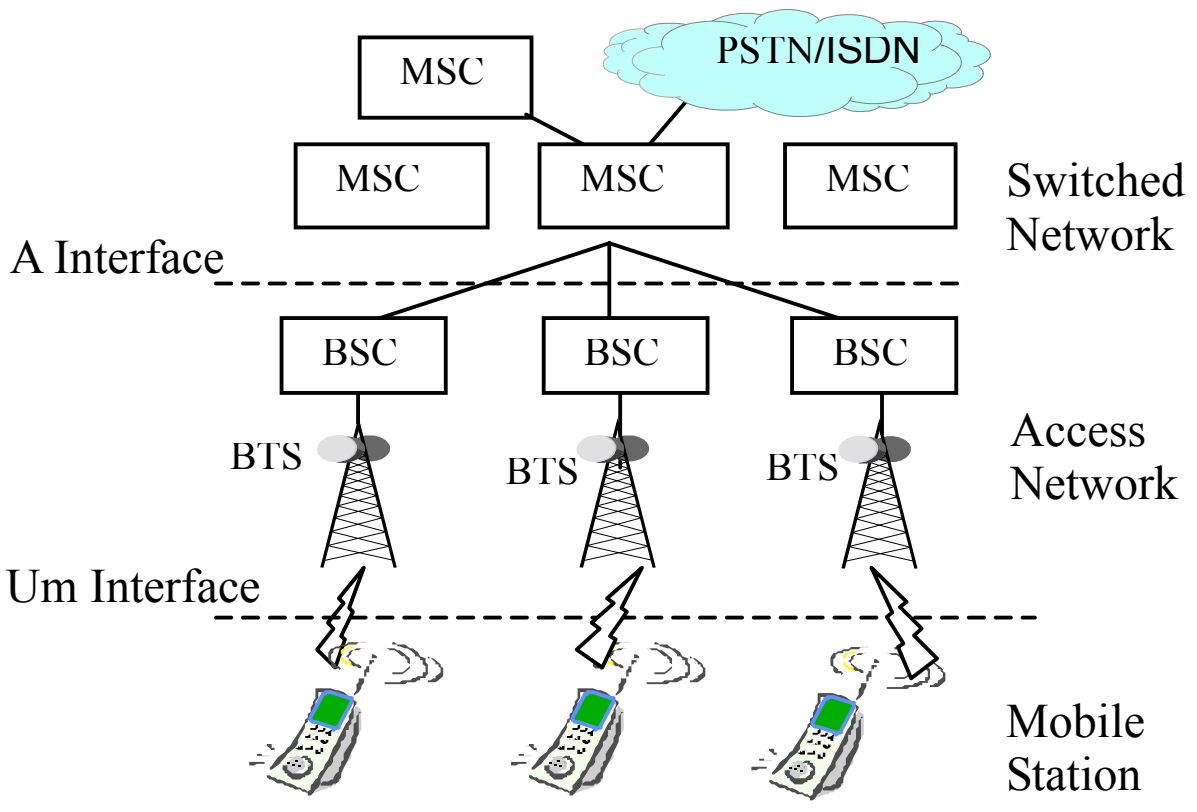

Figure 1. GSM Network

GSM is a second-generation, digital, TDMA-based, cellular stand with extensive coverage in Europe, including seamless coverage across European countries. Now GSM has been widely used in over 50 countries in the world.

A GSM network is composed of several functional entities, with defined functions and interfaces [24]. Figure 1 shows the layout of a generic GSM network. The GSM network can be divided into three broad parts: Mobile Station, Access Network and Switched Network..

- The Mobile Station is carried by the subscriber. The Base Station Subsystem controls the radio link with the Mobile Station.
- The Access Network consists of the Base Station Transceiver (BTS) and the Base Station Controller (BSC). The BSC provides the radio resource management, which serves the control and selection of appropriate radio channels to interconnect the MS and the Switched Network. The Access Network of GSM is also called the Base Station subsystem (BSS).
- The Switched Network, the main part of which is the Mobile services Switching Center, performs the switching of calls between the mobile and other fixed (PSTN/ISDN) or mobile network users, as well as management of mobile services, such as authentication.

The Mobile Station and the Base Station Subsystem (BSS) communicate across the Um interface, also known as the air interface or radio link. The Base Station Subsystem communicates with the Mobile service Switching Center across the A interface.

The basic interconnection of a mobile station to the fixed Public Switched Telephone Network (PSTN) is described in Figure 2. The communication between a mobile station and the base station subsystem (BSS) is through four types of channels: reverse control channel, forward control channel, reverse voice channel, and forward voice channel. These channels are implemented based on the Um interface, described above. A BSS will serve many mobiles in its coverage area. The BSS connects with an MSC through two types of connections, voice circuit and data, which can be carried over terrestrial or microwave links. An MSC may serve many BSSs.

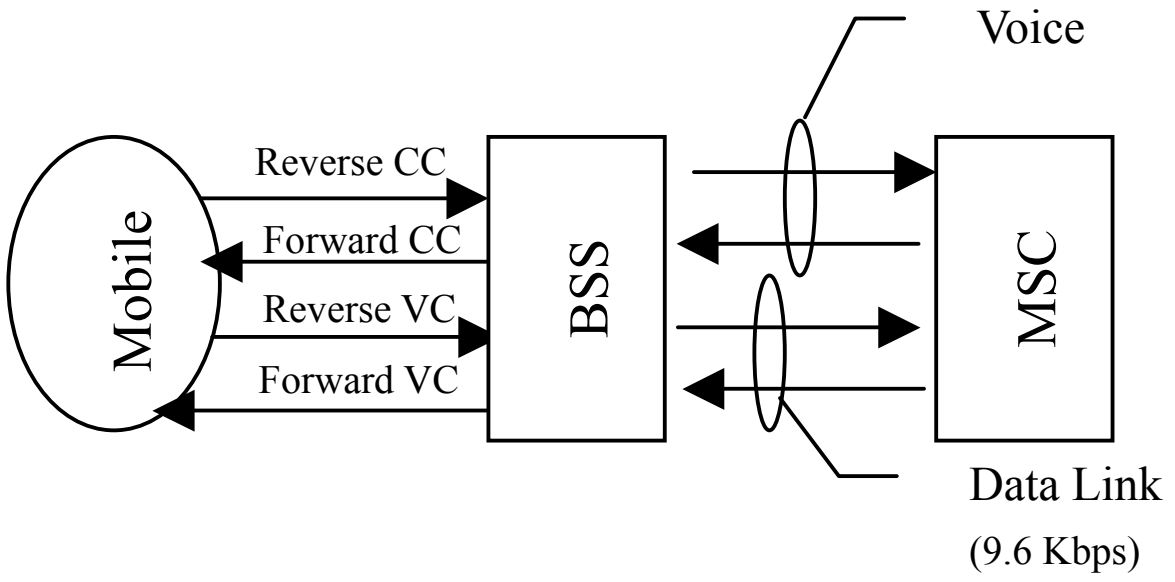

Figure 2. Mobile/BAS/MSC interconnection

## 3.2 GSM Network Configuration model

The concept of GSM networks topology can be considered as a tree. The first level of tree i.e. the root of tree is PLMN (Public land Mobile Network). The second level of tree is MSC (Mobile Service Switching Center) level, which corresponds to the Switched network in Figure 1. The third level of tree is BSC (Base Station Controller). The fourth level of tree is BTS level. The adjacent levels are connected by Data Link Channel, descried in Figure 2.

We can describe the concept of GSM networks topology illustrated in Figure 3 by 2-tuples: <object_class, parent_object_class> where object_class is the class of a network element and parent_object_class is the class of the upper level object of the network element.

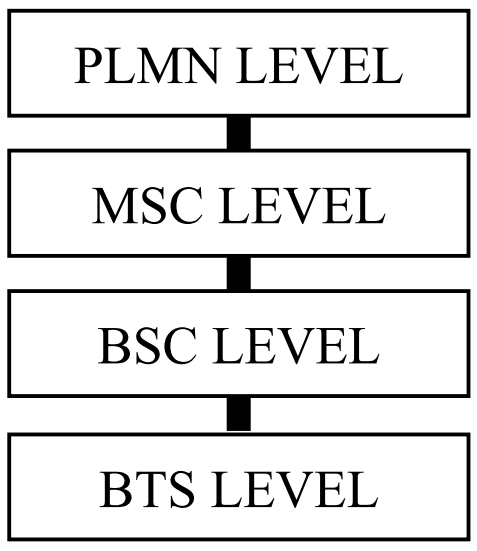

Figure 3. The Concept Structure of GSM

We can describe the structure of GSM networks in Figure 4 by 4-tuples <object_class, object_instance, parent_object_class, parent_object_instance> where object_class is the class of a network element, object_instance is the serial NO. of the network element, parent_object_class is the class of the upper level object of the network element, and parent_object_instance is the serial No. of the network element of the upper level object. The class of PLMN is defined as 0 and the instance of PLMN is defined as 0. Furthermore, PLMN hasn't parent class in GSM networks and it is the concept root of GSM networks. The classes of MSC, BSC, BTS are defined as 10,20 and 30, respectively. Note that the class of data circuit between MSC and BSC is defined as 12 and that between BSC and BTS is defined as 23. In the followings we'll give the features of the correlated alarms according to the configuration model of GSM.

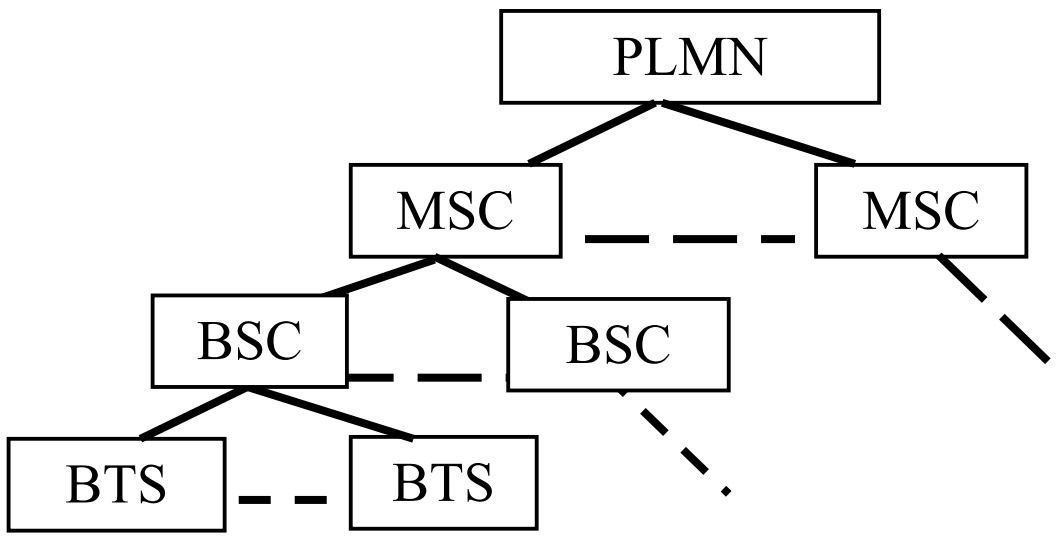

Figure 4. The structure of GSM

If the correlated alarms come from the same class of network element, then the degree of correlation among alarms is high. The alarms that are tightly correlated may indicate the mutual effect of components of network facility. If the correlated alarms come from the different class of network elements, then the degree of correlation among alarms will be lower than that of correlation among alarms coming from the same class of network elements. But the alarms that are loosely correlated may indicate the mutual effect of network elements in the whole networks, which will benefit planning and optimizing the whole networks. In what follows, we will give the definitions of Scope, Inter-correlated alarms and Intra-correlated alarms.

1.) **Scope** defines the area of alarms occurring in the networks. In this paper, we use the network element control area as the scope constraint in the GSM network. For example, if we specified MSC scope constraint, then the area includes the specified MSC, the BSCs that the MSC connects with and the BTSs that these BSCs connect with.
2.) **Inter-correlated alarms**: at least one of the correlated alarms belongs to the different class of network elements. By Inter-correlation constraint, we can make focus on discovering multi-level correlation relations among the MSC level, BSC level, and BTS level.
3.) **Intra-correlated alarms**: all of the alarms in the correlation rules must belong to the same class of network element i.e. BTS or BSC or MSC. By Intra-correlation constraint, we can discover the correlation relations in the same level of GSM network.

## 3.3 The Alarm Model [23]

**Definition 1. An alarm event**

An alarm event is defined as $E_i$=<$e_i$,$t_n$>, i, n=1,2,3,…, where $e_i$ is an alarm type and $t_n$ is the time of alarm occurring.

**Definition 2. An alarm type**

An alarm type is defined as $e_i$=<object_class,object_instance,alarm_num,desc>, i=1,2,3,…, where object_class is the serial NO. of object class, object_instance is the serial NO. of object instance, alarm_number is the NO. of alarm type and desc is the alarm information consisting of alarm priority and alarm description.

**Definition 3. An alarm tuple and its length**

1.) An alarm tuple is defined as $Q_i$=( ( $e_{k1}$, $e_{k2}$, …… , $e_{km}$), $t_i$ ), i, m=1,2,3,…; k1,k2,…,km=1,2,3,… , where '$e_{k1}$, $e_{k2}$,……, $e_{km}$' are the alarm types which concurrently occur at the time $t_i$ . An alarm tuple can be represented by an alarm event i.e. $Q_i$=(< $e_{k1}$, $t_i$>,< $e_{k2}$, $t_i$> ,……, <$e_{km}$, $t_i$>)=( $E_{k1}$, $E_{k2}$, ….. , $E_{km}$). If an alarm tuple $Q_i$ only contains one alarm type, then $Q_i$=(( $e_{k1}$), $t_i$)=( < $e_{k1}$, $t_i$>)= $E_{k1}$.
2.) The length of alarm tuple is $|Q_i|$=|( $e_{k1}$, $e_{k2}$, ……, $e_{km}$)|=m, which is the number of the alarm types contained in the alarm tuple.

**Definition 4. An alarm queue and its length**

1.) An alarm queue is defined as $S_{ij}$=<$Q_i$, $Q_{i+1}$,…, $Q_j$> i,j=1,2,3,…, where $t_i$< $t_{i+1}$ <…<$t_j$ . An alarm queue can be represented by an alarm event i.e. $S_{ij}$=<$Q_i$, $Q_{i+1}$,…,$Q_j$>=< ($E_{i1}$,.., $E_{ik}$), ($E_{i+1}$),…, ($E_{j1}$, …, $E_{jm}$)>= < ($E_{i1}$,…, $E_{ik}$), $E_{i+1}$,…, ($E_{j1}$, …, $E_{jm}$)>, i1,…,ik ,j1,…, jm=1,2,3,… .
2.) The length of alarm queue is defined as $|S_{ij}|$=|<$Q_i$, $Q_{i+1}$, …, $Q_j$>|=j-i+1, which is equal to the number of alarm tuple contained in alarm queue.

**Definition 5. An alarm type sequence and its length**

1.) An alarm type sequence is the m-tuple consisting of alarm types, which is denoted by $Seq_m$=< $e_{i1}$, $e_{i2}$,…,$e_{im}$> , where m=1,2,3, ⋯ ; i1,⋯,im=1,2,3,⋯ .
2.) Given alarm type sequence $Seq_m$=< $e_{i1}$, $e_{i2}$,…,$e_{im}$>,The length of alarm type sequence is defined as $|Seq_m|$=| < $e_{i1}$, $e_{i2}$,…,$e_{im}$>|=m.

**Definition 6**. Given an sequence $seq_m$ =< $e_{i1}$, $e_{i2}$,…,$e_{im}$ > and a transaction database DB, the times of the sequence $seq_m$ occurring in the DB are defined as

occur ($seq_m$, DB)=| the times of $seq_m$ occurring in DB|

**Definition 7.** Given an sequence $seq_m$ =< $e_{i1}$, $e_{i2}$,..,$e_{im}$ >, The support of $seq_m$ in database DB is defined as

$$\mathrm{support}(\mathrm{seq}_m, \mathrm{DB}) = \frac{\mathrm{occur}(\mathrm{seq}_m, \mathrm{DB})}{|\mathrm{DB}|}$$

Obviously, we have

$$\mathrm{occur}(\mathrm{seq}_m, \mathrm{DB}) = \mathrm{support}(\mathrm{seq}_m, \mathrm{DB}) \times |\mathrm{DB}|$$

**Definition 8.** Given two alarm type sequences: X, Y and an alarm viewing window $W_k$, the confidence of the alarm correlation rule X⇒Y is defined as

Conf(X⇒Y, $W_k$)=

|Support(XY,$W_k$)/Support(X, $W_k$)- Support(Y, $W_k$)|

**Definition 9.** A frequent alarm type sequence is defined as follows. In an alarm viewing window $W_k$, given that the minimum support of alarm type sequence is Mini_support, if the support of an alarm type sequence $seq_m$ is greater than Mini_support, then the alarm type sequence $seq_m$ is a frequent alarm type sequence.

**Definition 10. An alarm correlation rule**

$$e_{i1}, e_{i2},\ldots,e_{ij} \overset{\Delta t}{\Rightarrow} e_{ik}, e_{ik+1},\ldots, e_{im}$$

[conf=q%, supp=p% ,$W_k$]

After the alarm types $e_{i1}$, $e_{i2}$,…,$e_{ij}$ occur, in the interval of Δt, the probability of alarm type sequence <$e_{ik}$, $e_{ik+1}$,…, $e_{im}$> occurring is equal to q%.

# 4. Intelligent Search Algorithm with Model-based Constraints

The different classes of alarm correlation often have different support values in the alarm database. Some may be high, while others may be low. In many cases we only have interest in one class of alarm correlation. If we don't add the constraints to filter out these uninteresting classes of alarm correlation, the number of candidates in the algorithm will be very large. Thus the computing time of data mining will grow exponentially. But if we add the constraints to filter out these uninteresting correlated alarms, we can greatly enhance the performance of mining correlated alarms.

As mentioned in Section 3.2, the correlated alarms in the GSM network are divided into two groups: inter-correlated alarms and intra-correlated alarms. Because the inter-correlated alarms are loosely correlated, the support of inter-correlated alarms may be low. For the reason that the Intra-correlated alarms are tightly correlated, the values of support of Inter-correlated alarms may be high. If we have interest in inter-correlated alarms in the GSM networks, we can use the Inter-correlated constraint in the algorithm and don't need to worry about the computing time of data mining. It is the same thing for Intra-correlated constraint. Furthermore, we also use the Scope as the constraint of the algorithm to discover correlated alarms in the specified part of GSM networks.

**The Constraints definition based on the configuration model of GSM**

1.) **Scope constraint**: the area of the networks under the specified network element. Given the specified object_class i.e. sp_class, an alarm sequence set $C_m$ and an alarm sequence $seq_m$=<$e_{i1}$ ,$e_{i2}$,…,$e_{im}$> where $seq_m \in C_m$ · Scope constraint={ $seq_m \in C_m$ | $\forall$ $e_x$ in $seq_m$ ($e_x$.object_class ≥ sp_class) }.
2.) **Inter-correlated constraint**: at least two alarms belongs to the different classes of network elements. Given an alarm sequence set $C_m$ and an alarm sequence $seq_m$=<$e_{i1}$ ,$e_{i2}$,…,$e_{im}$> where $seq_m \in C_m$ · Inter-correlation constraint={ $seq_m \in C_m$ | $\exists$ $e_x$, $e_y$ in $seq_m$ ( $e_x$.object_class≠ $e_y$.object_class) }.
3.) **Intra-correlated constraint**: the alarms in the correlation rules must belong to the same class of network element i.e. BTS or BSC or MSC. Given an alarm sequence set $C_m$ and an alarm sequence $seq_m$=<$e_{i1}$ ,$e_{i2}$,…,$e_{im}$> where $seq_m \in C_m$.· Intra-correlation constraint={ $seq_m \in C_m$ | $\forall e_x$, $e_y$ in $seq_m$ ($e_x$.object_class= $e_y$.object_class) }.

After the above analysis, we propose the intelligent search of correlated alarms with model-based constraint, called SMC (Search with Model-based Constraints),which consists of Algorithm 1, Algorithm 2, Algorithm 3 and Algorithm 4.

The Algorithm 1 is the main part of the SMC algorithm. Algorithm 1 is based on the framework of the algorithms [6,7,8] and composed of Algorithm 2 and Algorithm 3. The Algorithm 2 is the Robust_search algorithm in [23], which search the times of an alarm type sequence occurring in the given alarm queue. Algorithm 3 generates the frequent alarm type sequences $F_ALARM_{m+1}$ from the candidate alarm sequences $C_m$.

The constraints, i.e. Scope constraint, Inter-correlated constraint and Intra-correlated constraint are in line 8 of Algorithm 1. We use these constraints to prune the candidates that don't satisfy the constraints mentioned above. Thus, we can reduce the number of frequent alarm type sequences and so reduce the computing time of the algorithm. By use of the constraints, we could control the process of mining alarms and focus on mining the sequences, which are interesting to us.

The Algorithm 4 generates the correlated rules from the frequent alarms. The interesting measure of correlation rules in the frequent pattern mining was first studied by Brin et al.[21]. Later K. M. Ahmed et al. [22] improved the interesting measure that was presented by Brin et al. we use the interesting measure i.e. |P(XY)/P(X)-(Y)| of correlation rule X⇒Y in [22] as the one in Algorithm 4.

### ■ Algorithm 1

Input: alarm queue ($S_{ij}$, $W_k$)
Output: t frequent alarm sequence set: $F_ALARM_m$

1. compute $C_1$:={ α | α∈ $F_ALARM_1$};
2. m:=1;
3. while $C_m \neq \Phi$ do
4. begin
5. For all α∈ $C_m$ , Search alarm queue $S_{ij}$ to find support(α, $W_k$); /*Algorithm 2 */
6. $F_ALARM_m$={ α∈ $C_m$| support(α, $W_k$)≥ min_support ∧ ( ( α∈ **Inter-correlated-constraint** ) || ( α∈ **Intra-correlated-constraint** ) ∧ ( α∈ **Scope-constraint** ));
7. Generate Candidate $C_{m+1}$ from $F_ALARM_m$; /* Algorithm 3 */
8. m=m+1;
9. end.
10. for all m , output $F_ALARM_m$;

## ■ Agorithm 3

Input: frequent alarm sequence set $F_ALARM_m$.

Output: frequent alarm sequence Candidate set $C_{m+1}$.

1. $C_{m+1}:=\Phi$;
2. For $\alpha,\beta \in F_ALARM_m$ and $\alpha \neq \beta$ and $\alpha=< e_{i1}, e_{i2},\ldots, e_{im}>$, $\beta=< e_{i1}`, e_{i2}`,\ldots, e_{im}`>$
3. begin
4. If ($e_{i2}= e_{i1}` \cap e_{i3}= e_{i2}` \cap,\ldots,\cap e_{im}= e_{i\,m-1}`$ ) then begin
5. generate alarm sequence $\gamma=< e_{i1}, e_{i2},\ldots, e_{im}, e_{im}`>$; /* Candidate generate */
6. $C_{m+1}:= \{\gamma \mid$ For all $L \subseteq \gamma$ and $|L|=m$ ,we have $L \in F_ALARM_m \}$; /*Pruning Candidate */
7. end
8. end

## ■ Algorithm 4

Input: Frequent alarm sequence set $F_ALARM_m$

Output: output the correlation rules $\beta \rightarrow (\alpha-\beta)$ and confidence $|P(\alpha)/P(\beta)-P(\alpha-\beta)|$

1. for all $\alpha \in F_ALARM_m$ do /* generate correlation rules */
2. for all $\beta \subseteq \alpha$ do
3. if$|P(\alpha)/P(\beta)-P(\alpha-\beta)| \geq min_conf$ then
4. begin
5. generate the rule $\beta \rightarrow (\alpha-\beta)$ with
6. confidence $|P(\alpha)/P(\beta)-P(\alpha-\beta)|$ ;
7. end

# 5. Experiments

We conducted a set of experiments to test the performance of SMC (Search with Model-based Constraints) algorithm. The experiments were on the DELL PC Server with 2 CPU Pentium II, CPU MHz 397.952211, Memory 512M and SCSI Disk 16G. The Operating system on the server is Red Hat Linux version 6.0.

The data in experiments are the alarms in GSM Networks, which contain 181 alarm types and 90k alarm events. The time of alarm events range from 2001-03-15-00 to 2001-03-19-23. In Figure 5, Figure 6 and Figure 7 Intra-correlated constraint, Intra-correlated constraint and no constraints are denoted by **Intra, Inter** and **Nocons** respectively. In experiments, we use the minimum times of alarm sequence occurring as the Minimum support.

First we give the definition of average times of alarm sequence occurring. *Average times* of alarm sequence occurring is equal to the sum of times of all alarm sequences divided by the number of alarm sequences.

.

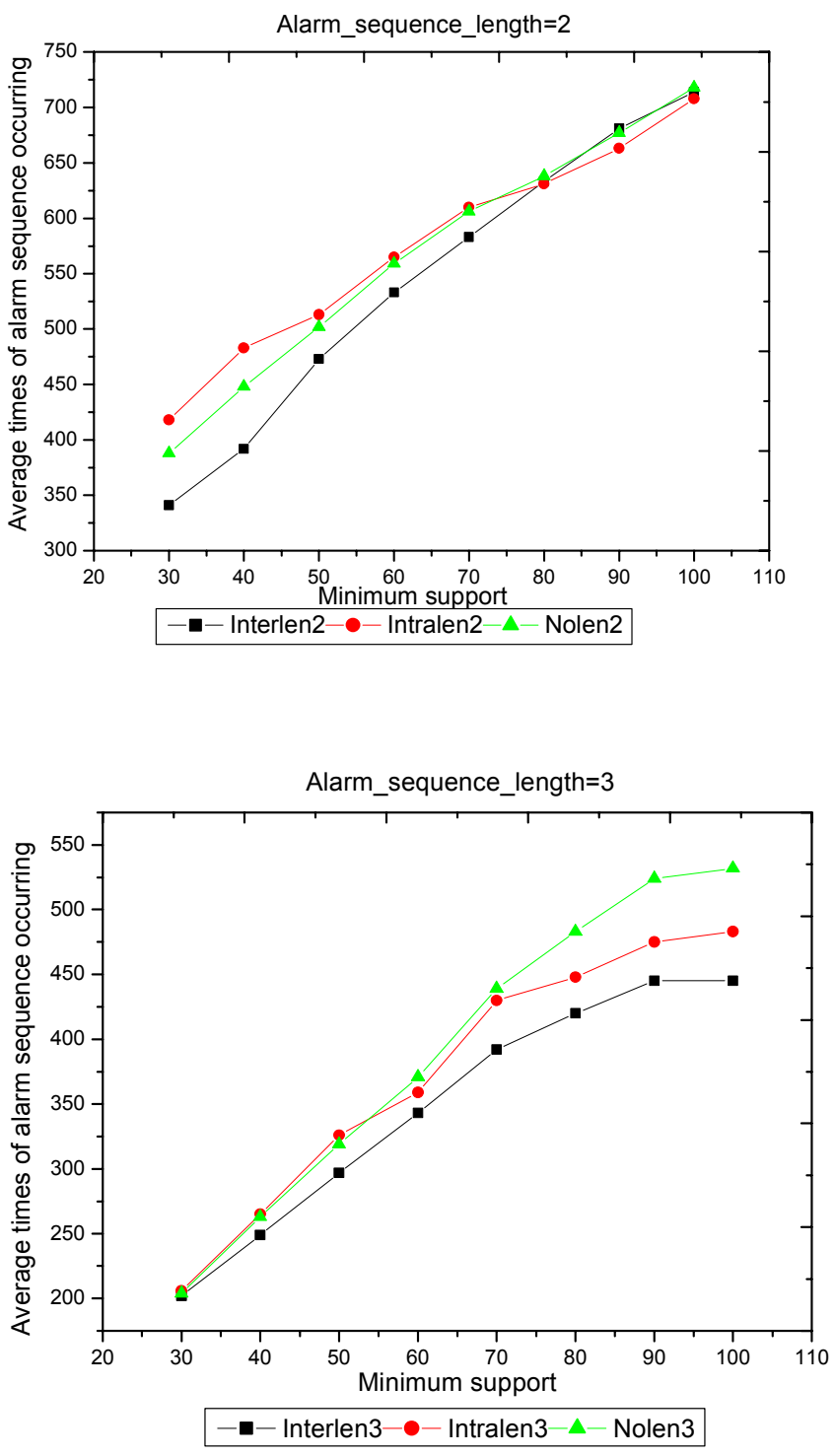

Figure 5 Average times of alarm sequences of Inter, Intra, no constraints

The graph of alarm_sequence_len=2 indicates that the broken lines of Intra, Inter and Nocons are to converge with the increment of Minimum support, while the graph of alarm_sequence_len=3 indicates that the broken lines of Intra, Inter and Nocons are to diverge with the increment of Mini_support. In sum, the average times of alarm sequence occurring increases with the increment of Minimum support. The average times of Intra-correlated constraint alarms are larger than these of Inter-correlated constraint alarms with the same Minimum support, which indicates that Inter-correlated alarms are more correlated than Intra-correlated alarms.

From Figure 6, it is obvious to see that with the increment of Minimum support, the executing time of algorithms will be smaller than before. The algorithm with Inter-correlated constraint and Intra-correlated constraint are two times faster than that with no constraints at the same Minimum support. The algorithm with Inter-correlated constraint is two times faster than that with Intra-correlated constraint at the same Minimum support. From Figure 7, The number of alarm sequences discovered by the algorithm with no constraints are about three times of the one discovered by that with Inter-correlated constraint and about two times of the one discovered by that with Intra-correlated constraint.

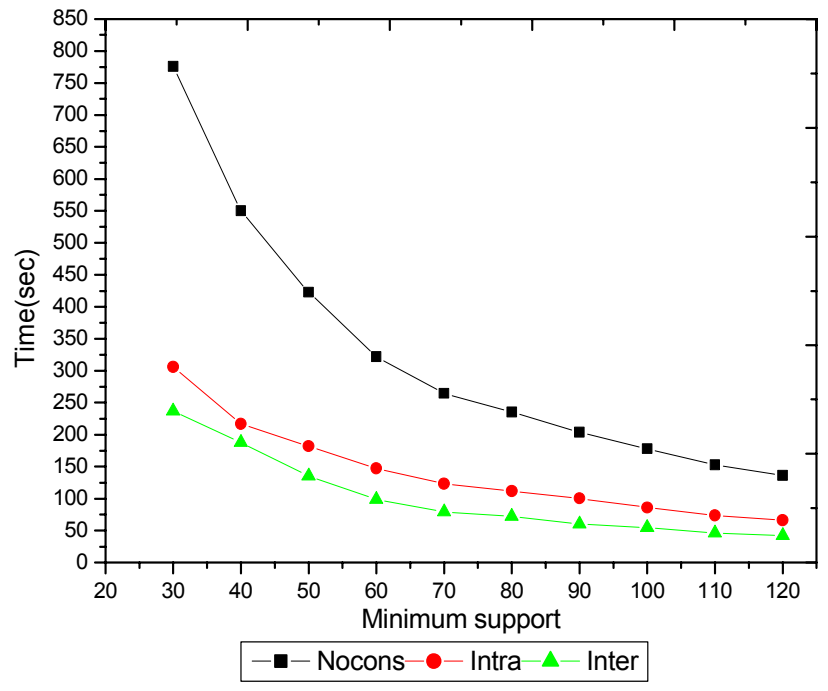

Figure 6 Execution time

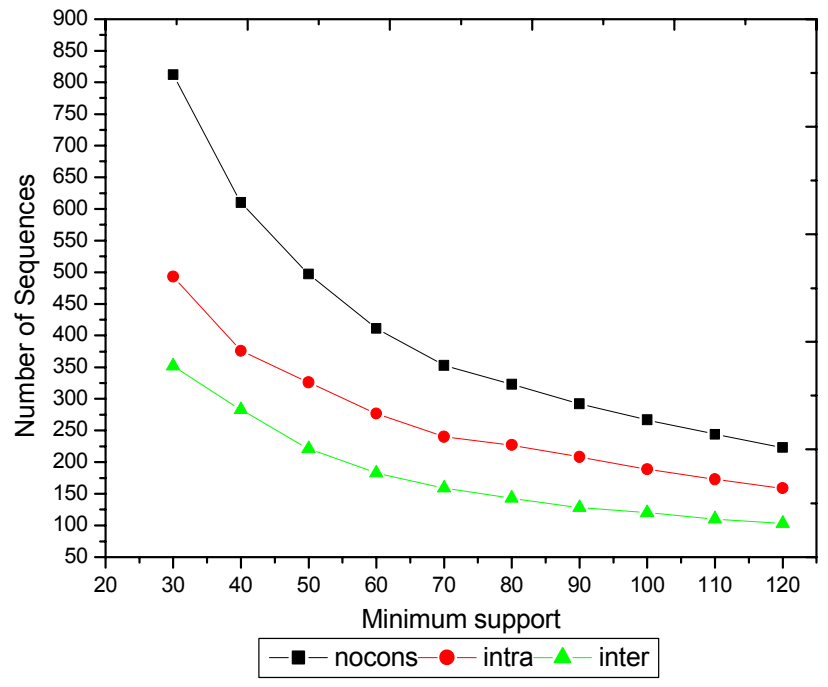

Figure 7 Number of sequences

# 6. Conclusion

We presented a new algorithm for intelligent search of correlated alarms with model-based constraints. In the algorithm we use three kinds of constraints i.e. Scope constraint, Inter-correlated constraint and Intra-correlated constraint. The experiments shows that the algorithm with the Inter-correlated constraint and Intra-correlated constraint are about two times faster than the one with no constraints. From the result of experiments, we can see that the Inter-correlated alarms are more correlated than the Intra-correlated alarms. The constraints for alarm correlation are not limited to the three kinds of ones above. According to the features of the thing that we want to discover, we can design new kinds of constraints for the intelligent search of correlated alarms.

# Acknowledgements

Thanks Professor LI Wei for the choice of subject and guidance of methodology. Thanks for the suggestions from Professor SUI YueFei of Chinese Academy Sciences. Thanks for Nan Liu, Zhi Cui and Xinzhang Li of Beijing Mobile Telecom. The author would like to thank Xin Gao, Peng Cheng, Gang Zhou , the other member of the NLSDE lab.